\documentclass[twocolumn,english,rmp, a4paper]{revtex4}
\usepackage[T1]{fontenc}
\usepackage[latin9]{inputenc}
\usepackage{geometry}
\usepackage{color}
\usepackage{graphicx}
\usepackage{amssymb}

\makeatletter

\providecommand{\boldsymbol}[1]{\mbox{\boldmath $#1$}}

\providecommand{\tabularnewline}{\\}

\usepackage{geometry}

\usepackage{babel}
\makeatother

\begin{document}

\title{Understanding adsorption of hydrogen atoms on graphene}

\author{Simone Casolo}

\address{\textcolor{black}{University of Oslo, Department of Physics, P.O.Box
1048 Blindern, NO-0316 Oslo, Norway}. On leave from Dept. of Physical
Chemistry and Electrochemistry, University of Milan }

\author{Ole Martin Løvvik}

\address{\textcolor{black}{University of Oslo, Department of Physics, P.O.Box
1048 Blindern, NO-0316 Oslo, Norway and SINTEF Materials and Chemistry,
Forskningsvn 1, NO-0314 Oslo, Norway }}

\author{Rocco Martinazzo }

\address{Dept. of Physical Chemistry and Electrochemistry and CIMAINA, University
of Milan, V. Golgi 19, 20133 Milan, Italy}

\email{rocco.martinazzo@unimi.it}

\author{Gian Franco Tantardini}

\address{Dept. of Physical Chemistry and Electrochemistry and CIMAINA, University
of Milan, V. Golgi 19, 20133 Milan, Italy and ISTM Institute for Molecular
Science and Technology, V. Golgi 19, 20133 Milan, Italy }

\begin{abstract}
Adsorption of hydrogen atoms on a single graphite sheet (graphene)
has been investigated by first-principles electronic structure means,
employing plane-wave based, periodic density functional theory. A
reasonably large 5x5 surface unit cell has been employed to study
single and multiple adsorption of H atoms. Binding and barrier energies
for sequential sticking have been computed for a number of configurations
involving adsorption on top of carbon atoms. We find that binding
energies per atom range from $\sim0.8$ eV to $\sim1.9$ eV, with
barriers to sticking in the range $0.0-0.2$ eV. In addition, depending
on the number and location of adsorbed hydrogen atoms, we find that
magnetic structures may form in which spin density localizes on a
$\sqrt{3}\mbox{x}\sqrt{3}\mbox{R}30^{\circ}$ sublattice, and that
binding (barrier) energies for sequential adsorption increase (decrease)
linearly with the site-integrated magnetization. These results can
be rationalized with the help of the valence-bond resonance theory
of planar $\pi$ conjugated systems, and suggest that preferential
sticking due to barrierless adsorption is limited to formation of
hydrogen pairs. 
\end{abstract}
\maketitle

\section{Introduction }

Recent years have witnessed an ever growing interest in carbon-based
materials. Carbon, being a small atom with a half-filled shell, is
able to mix its valence $s$ and $p$ orbitals to various degrees,
thereby forming the building block for extended structures of incredibly
different electronic, magnetic and mechanical properties. Among them,
those formed by $sp^{2}$ C atoms have attracted much attention in
the last few years. They can be collective termed as graphitic compounds
and comprise graphite, carbon nanotubes, fullerenes, Polycyclic Aromatic
Hydrocarbons (PAHs), and recently graphene (the one-atom thick layer
of graphite) and graphene nanoribbons (GNRs). In particular, the revolutionary
(and embarrassing simple) fabrication of graphene \citep{novoselov04}
has opened the way for a wealth of studies in both fundamental and
applied science. New, extraordinary properties have become available
to material design since its isolation. Indeed, even though they have
been known since the first theoretical analysis by Wallace \citep{wallace47},
it was only the experimental observation of the existence of one-atom
tick layer of graphite that triggered much of the current interest.
In particular, one of the most interesting aspects of graphene is
that it presents low energy excitations as massless, chiral, Dirac
fermions mimicking the physics of quantum electrodynamics \citep{novoselov05,zhang05,castroneto08}. 

In this context, adsorption of hydrogen atoms on graphene and GNRs
can be used to tailor electronic and magnetic properties, as already
suggested for other `defects', with the advantage of being much easier
to realize than e.g. vacancies. In addition, interaction of hydrogen
atoms with graphitic compounds has been playing an important role
in a number of fields as diverse as nuclear fusion \citep{parrinello01,mayer01},
hydrogen storage \citep{Schlapbach2001} and interstellar chemistry
\citep{hartquist-book}. 

In material design for hydrogen storage, several carbon based structures
has been proposed as candidates \citep{Schlapbach2001}, in particular
in connection with the spillover effect following embedding of metallic
nanoparticles. Though these materials are in practice still far from
the weight percent target stated by the US department of Energy, they
remain a cheap and safe alternative, and a deeper understanding of
the mechanisms underlying adsorption may lead in future to a more
efficient material design. 

In interstellar chemistry hydrogen-graphite and hydrogen-PAHs systems
have become realistic models to investigate molecular hydrogen formation
in the interstellar medium (ISM). There are still open questions in
this context since, in spite of continuous destruction by UV radiation
and cosmic rays, $H_{2}$ is the most abundant molecule of the ISM.
It is now widely accepted that $H_{2}$ can only form on the surface
of interstellar dust grains and particles \citep{Gould-Salpeter1963,Hollenbach1970,Hollenbach1971},
which -with the exception of cold, dense molecular clouds- are either
carbon-coated silicate grains or carbonaceous particles or large PAHs
\citep{Greenberg2002,Williams2002,Draine2003}. This finding has stimulated
a number of theoretical \citep{Jeloaica1999,Sha&Jackson2002,Sha2002,Zecho2002,Sha2005,Morisset2004,Allouche2005,Morisset2005,Martinazzo2005b,Allouche2006,Kerwin2006,Martinazzo2006a,Martinazzo2006b,Bonfanti2006,Cuppen2008,jackson08}
and experimental \citep{Zecho2002,Guttler2004a,Zecho2004,Guttler2004,Andree06,Hornekaer2006,Hornekaer2006a,Baouche2006,Creighan2006,price07,Hornekaer2007}
studies on hydrogen graphitic systems aimed at elucidating the possible
reaction pathways leading ultimately to molecule formation. 

One interesting finding of these studies is the tendency of hydrogen
atoms to cluster at all but very low coverage conditions \citep{Hornekaer2006,Hornekaer2006a,Andree06,Hornekaer2007}.
New mechanisms for hydrogen sticking \citep{Hornekaer2006a} and new
recombination pathways \citep{Hornekaer2006} have been proposed,
based on the now common agreement that the presence of one or more
adsorbate atoms strongly influences subsequent adsorption. It is clear
that such an influence can only result as a consequence of a substrate-mediated
interaction which makes use of the unusual electronic properties of
graphitic compounds, but at present a comprehensive model for multiple
chemisorption is still missing. 

In this work we present first principles calculations of single and
multiple adsorption of hydrogen atoms on a graphene sheet, used as
a model graphitic material, with the aim of understanding the relationship
between the substrate electronic properties and the stability of various
cluster configurations. This work parallels analogous investigations
of defects in graphene and GNRs \citep{Pereira2006,Yazyev2007,pisani08,pereira08a,palacios08,yazyev08d}.
Indeed, they all share the disappearance of one or more carbon $p$
orbitals from the $\pi-\pi^{*}$ band system, a fact which may lead
to the appearance of magnetic textures and introduce site-specific
dependence on the chemical properties. Complementing previous investigations,
however, we show how the simple $\pi$ resonance chemical model helps
in rationalizing the findings. A parallel work on different graphitic
substrates (PAHs) will follow shortly \citep{bonfanti08}. 

The paper is organized as follows. Details of our first-principles
calculations are given in Section \ref{sec:Computational-methods},
and their results in Section \ref{sec:Results}, where we analyze
adsorption of a single H atom and briefly introduce the chemical model
(Section \ref{sub:Single-atom-adsorption}), we consider formation
of pairs (Section \ref{sub:Secondary-adsorption}) and formation of
three- and four- atom clusters (Section \ref{sub:Further-adsorptions}).
We summarize and conclude in Section  \ref{sec:Summary-and-Conclusions}.

\section{\label{sec:Computational-methods}Computational methods}

Periodic density functional theory as implemented in the Vienna \emph{Ab}
\emph{initio} Simulation Package suite (VASP) \citep{VASP1,VASP2,VASP3,VASP4}
has been used in all the calculations. The projector-augmented wave
method within the frozen core approximation has been used to describe
the electron-core interaction \citep{PAW1,PAW2}, with a Perdew-Burke-Ernzerhof
(PBE) \citep{PBE1,PBE2} functional within the generalized gradient
approximation (GGA). Due to the crucial role that spin plays in this
system all our calculations have been performed in a spin unrestricted
framework. 

All calculations have used an energy cutoff of 500 eV and a 6x6x1
$\Gamma$-centered \emph{k}-points mesh to span the electron density,
in a way to include all the special points of the cell. The linear
tetrahedron method with Bl\"ochl corrections is used \citep{LinTetrah}
together with a 0.2 eV smearing. All the atomic positions have been
fully relaxed until the Hellmann-Feynman forces dropped below $10^{-2}\ \mbox{eV}$\AA$^{-1}$,
while convergence of the electronic structures has been ensured by
forcing the energy difference in the self consistent cycle to be below
$10^{-6}\mbox{eV}$ , with the exception of energy barriers determination
where the thresholds were $10^{-5}\ \mbox{eV}$ for electrons and
$10^{-4}\ \mbox{eV}$ for ions. We have checked that both setups give
the same results within a meV accuracy. 

The slab-supercell considered has been carefully tested and a 20 \AA \  
vacuum along the \emph{c }axis has been adopted to ensure no reciprocal
interaction between periodical images. We find that using the above
settings the interaction between two adjacent graphite layers is $\sim$2
meV, largely within the intrinsic DFT error. This result is in agreement
with literature data \citep{Hasegawa2004,Rydberg2003}. For this reason
a single graphene sheet can also model the Bernal (0001) graphite
surface, at least as long as chemical interactions are of concern. 

The cell size on the surface plane is a fundamental parameter for
these calculations, since we have found that chemisorption energies
strongly depend on the coverage (see below). We choose to use a reasonably
large 5x5 cell in order to get some tens of meV accuracy while keeping
the computational cost as low as possible. Even with this size, however,
the possibility of interactions between images has always to be taken
into account when rationalizing the data. %
\begin{table}
\begin{centering}
\begin{tabular}{|c|c|cc|cc|}
\hline 
\textcolor{black}{unit cell} & \textcolor{black}{$\theta$ / ML } & \textcolor{black}{d$_{puck}$ /\AA} &  & \textcolor{black}{E$_{chem}$ / eV} & \tabularnewline
\hline
 &  & \textcolor{black}{this work} & \textcolor{black}{others} & \textcolor{black}{this work } & \textcolor{black}{others}\tabularnewline
\hline 
\textcolor{black}{2x2} & \textcolor{black}{0.125} & \textcolor{black}{0.36} & \textcolor{black}{0.36$^{1}$} & \textcolor{black}{0.75} & \textcolor{black}{0.67$^{1}$}\tabularnewline
\hline 
\textcolor{black}{3x3} & \textcolor{black}{0.062} & \textcolor{black}{0.42} & \textcolor{black}{0.41$^{2}$} & \textcolor{black}{0.77} & \textcolor{black}{0.76$^{2}$}\tabularnewline
\hline 
\textcolor{black}{cluster} & \textcolor{black}{0.045} & \textcolor{black}{-} & \textcolor{black}{0.57$^{3}$} & \textcolor{black}{-} & \textcolor{black}{0.76$^{3}$}\tabularnewline
\hline 
\textcolor{black}{4x4} & \textcolor{black}{0.031} & \textcolor{black}{0.48} & \textcolor{black}{-} & \textcolor{black}{0.79} & \textcolor{black}{0.76$^{4}$, 0.85$^{5}$}\tabularnewline
\hline 
\textcolor{black}{5x5} & \textcolor{black}{0.020} & \textcolor{black}{0.59} & \textcolor{black}{-} & \textcolor{black}{0.84} & \textcolor{black}{0.71$^{6}$, 0.82$^{7}$}\tabularnewline
\hline 
\textcolor{black}{8x8} & \textcolor{black}{0.008} & \textcolor{black}{-} & \textcolor{black}{-} & \textcolor{black}{-} & \textcolor{black}{0.87$^{8}$}\tabularnewline
\hline
\end{tabular}
\par\end{centering}

\caption{Chemisorption energy (E$_{chem}$) and equilibrium height of the C
atom above the surface (d$_{puck}$) for H adsorption on top of a
C atom, for a number of surface unit cells, corresponding to different
coverages $\theta$. Ref. 1 \citet{Sha&Jackson2002}, Ref. 2 \citet{Kerwin2008},
Ref. 3 \citet{Ferro2003}, Ref. 4 \citet{Duplock2004}, Ref. 5 \citet{Hornekaer2006a},
Ref. 6 \citet{Roman2007}, Ref. 7 \citet{ChenSpillover}, Ref. 8 \citet{Lethinen2004}.
\label{tab:single-atom}}

\end{table}

\section{\label{sec:Results}Results}

\subsection{\label{sub:Single-atom-adsorption}Single atom adsorption}

Chemisorption of single H atoms on graphite has long been studied
since the works of \citet{Jeloaica1999} and \citet{Sha&Jackson2002},
who first predicted surface reconstruction upon sticking. Such a reconstruction,
i.e. the puckering of the carbon atom beneath the adsorbed hydrogen
atom, occurs as a consequence of  $sp^{2}-sp^{3}$ rehybridization
of the valence C orbitals needed to form the $\sigma$ CH bond. Since
this electronic/nuclear rearrangement causes the appearance of an
energy barrier $\sim$0.2 eV high, sticking of hydrogen atoms turns
out to be a thermally activated process which hardly occurs at and
below room temperature \citep{Kerwin2008}. 

As already said in Section II, we have re-considered adsorption of
single hydrogen atoms for different sizes of the surface unit cell.
We have found that both the binding energy and the puckering height
are strongly affected by the size of the unit cell (see Table \ref{tab:single-atom}),
and even the results of the 5x5 cell turns out to be in error of about
$\sim$30 meV with respect to the isolated atom limit estimated by
the calculation at $0.008$ ML coverage \citep{Lethinen2004}. In
particular, we have found that some cautions is needed in comparing
the height of the carbon atom involved in the bond since constraining
the neighboring carbon atoms in geometry optimization may lead to
considerable surface strain. 

Despite this, we have consistently used the 5x5 cell in studying multiple
adsorption of hydrogen atoms. Indeed, this size allowed us to investigate
a number of stable configurations involving two, three and four adsorbed
H atoms, along with the barrier to their formation, with the same
set-up described in Section II. Interactions between images do indeed
occur for some configurations but, as we show below, this does not
prevent us to get a clear picture of the adsorption processes we are
interested in. %
\begin{figure}
\begin{centering}
\includegraphics[clip,width=0.7\columnwidth]{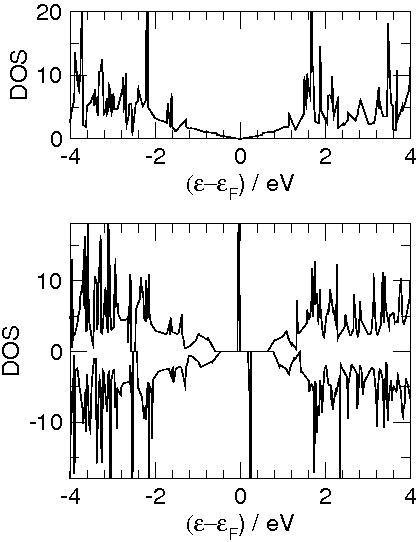}
\par\end{centering}

\caption{\label{fig:1H-DOS}Top panel: total density of states for graphene.
Bottom panel: density of states for spin-up (positive values) and
spin-down (negative values) components in a 5x5 H layer on graphene. }
\label{fig:curva1H}
\end{figure}

In agreement with previous studies we find that hydrogen adsorption
can only occur if the substrate is allowed to relax. Without relaxation
the adsorption curves on different surface sites are repulsive, and
only a metastable minimum is found for the atop position \citep{Sha&Jackson2002}.
Surface relaxation requires about $0.8-0.9$ eV and results in the
outward motion of the carbon atom forming the CH bond (see Table \ref{tab:single-atom}
and \citet{Jeloaica1999,Sha&Jackson2002}). 

\begin{figure}
\begin{centering}
\includegraphics[clip,width=1\columnwidth]{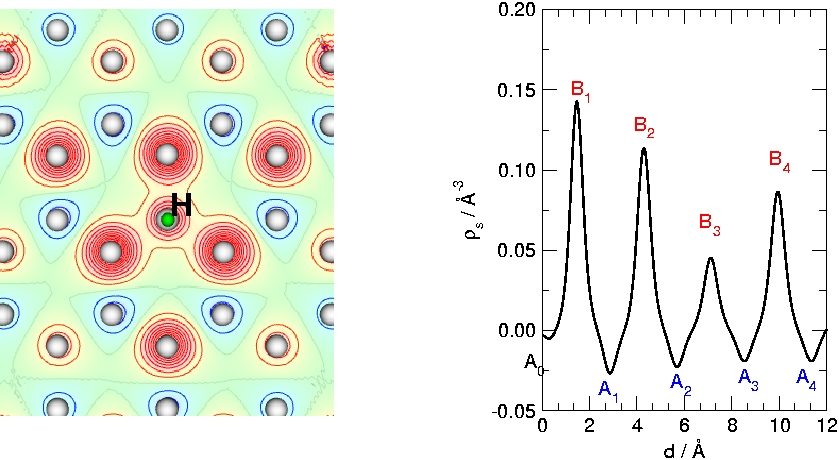}
\par\end{centering}

\caption{\label{fig:SpinDensity}Spin density 0.47 \AA above the graphene
surface after adsorption of a hydrogen atom. Left: contour map with
red/blue lines for spin-up/spin-down excess respectively. Right: spin-density
at the same height as on the left panel, along a path joining the
C atoms (for the labels see Fig.\ref{fig: double adsorption sites}). }

\end{figure}

In addition, we have investigated the electronic substrate properties
of the resulting hydrogenated graphene, in order to get hints for
understanding the adsorption process of additional atoms. In Fig.
\ref{fig:1H-DOS} we show the Density of States (DOS) of the 5x5 H-graphene
equilibrium structure (bottom panel), compared to that of clean graphene
(top panel). It is evident from the figure that hydrogen adsorption
causes the appearance of a double peak in the DOS, symmetrically placed
around the Fermi level. This is in agreement with rigorous results
that can be obtained in tight binding theory for \emph{bipartite}
lattices. Indeed\emph{, }\citet{Inui1994} have shown that for a bipartite
lattice with $n_{A}$ A lattice sites and $n_{B}$ B lattice sites
a sufficient condition for the existence of mid-gap states is a lattice
imbalance ($n_{A}\ne n_{B}$). In particular, there exist $n_{I}=|n_{A}-n_{B}|$
mid-gap states with \emph{vanishing} wavefunction on the minority
lattice sites. In H-graphene a lattice imbalance results as a consequence
of the bond with the H atom which makes one of $p$ orbitals no longer
available for taking part to the $\pi-\pi^{*}$ band system. There
is one mid-gap state for each spin species, and the degeneracy is
lifted if exchange-correlation effects are taken into account, as
shown in Fig.\ref{fig:1H-DOS} for our DFT results. This state has
been mapped out in Fig.\ref{fig:SpinDensity} (left panel), where
we report a contour map of the spin density at a constant height 0.47
\AA ~above the surface. It is clear from the figure that if adsorption
occurs on a A lattice site the spin-density (due to mainly to the
above mid-gap state) localizes on B lattice sites. The latter now
contain most of the 1 $\mu_{B}$ magnetization ($\mu_{B}=$Bohr magneton)
previously carried by the H atom species, and a slight spin-down excess
on A sites results as a consequence of the spin-polarization of the
lower lying states. This is made clearer in the right panel of Fig.
\ref{fig:SpinDensity} where we report the spin-density at the same
height above the surface as before along a rectilinear path joining
a number of C atom sites away from the adsorption site (see Fig.\ref{fig: double adsorption sites}
for the labels). Note that the spin-density decays only slowly with
the distance from the adsorption site, in agreement with theoretical
results that suggest that in the case of two dimensional graphene
this decay corresponds to a non-normalizable state with a $1/r$ tail
(in contrast to non-zero gap substrates such as armchair nanoribbons
where mid-gap states are normalizable)\citep{Pereira2006}. With our
unit cell the effect of the interaction with the images is already
evident at rather short distances, but as we show below, this effect
has no influence on the interpretation of the results. Note also that
this spin pattern is common to other `defects' (e.g. vacancies, voids
and edges) which have been known for some time to strongly modify
the electronic properties of graphene and graphene-like structures,
and to (possibly) produce long-range ordered magnetic structures \citep{Mizes1989,Ruffieaux2000,Kusakabe03,Lethinen2004,Pereira2006,Yazyev2007,Jiang,yazyev08a,pisani08,pereira08a,yazyev08c,palacios08,yazyev08d}.
In particular, in a recent, comprehensive study, Palacios \emph{et
al.,} using a mean-field Hubbard model for graphene, have clarified
the appearance of magnetic textures associated to vacancies and predicted
the emergence of magnetic order \citep{palacios08}. Their model also
suits well to `defects' such as the presence of the adsorbed hydrogen
atoms. 

\begin{figure}
\begin{center}
\includegraphics[clip,width=0.9\columnwidth]{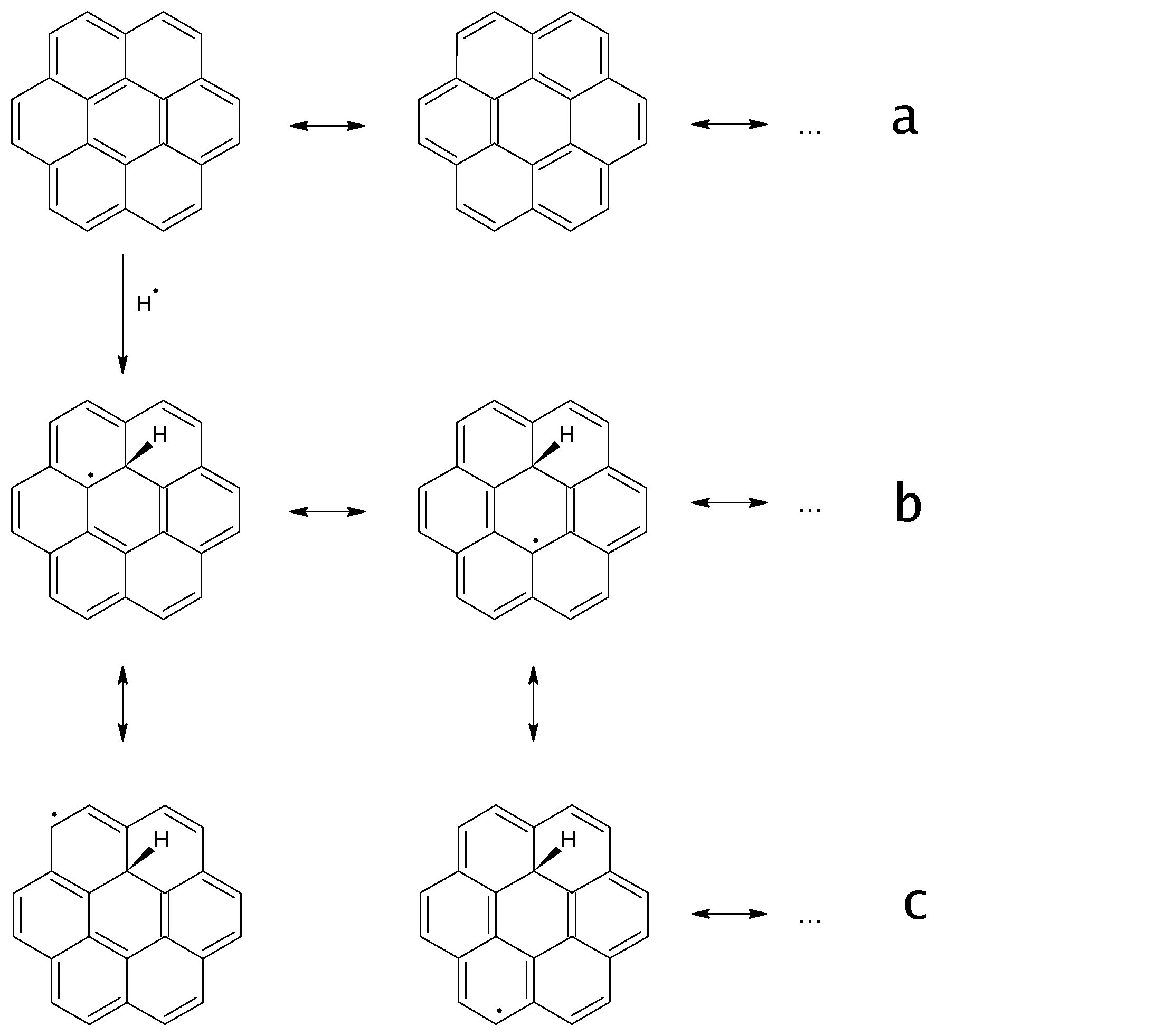}
\end{center}

\caption{\label{fig:coronene-model}(a) The $\pi$ resonating chemical model
for a graphenic surrogate (coronene). (b), (c) Spin-alternation after
hydrogen adsorption.}

\end{figure}

From a chemical point of view the above spin pattern (and the resulting
magnetic properties) arise from the `spin-alternation' typical of
$\pi$ conjugated compounds. This behavior is easily understood in
terms of resonant chemical structures, such as those shown in Fig.\ref{fig:coronene-model}
for a coronene molecule. In this and analogous Polycyclic Aromatic
Hydrocarbons (PAHs), the $\pi$ electron system can be described as
a resonant combination of conventional, alternated double bond structures,
like the one shown in the upper panel of Fig.\ref{fig:coronene-model}
(see \emph{a}). Once a hydrogen atom has been adsorbed on the surface,
an unpaired electron is left on one of the neighboring C atoms (\emph{b},
left panel), which can subsequently move in each of the carbon atoms
belonging to a sublattice $\sqrt{3}\mbox{x}\sqrt{3}\mbox{R}30^{\circ}$
by `bond-switching' (see\emph{ b,c}). Spin-alternation arises from
the `resonant' behavior of an unpaired electron in $\alpha$ position
(the nearest neighbor one) with respect to a double bond: such `resonance'
can be naively viewed as the spin re-coupling of the unpaired electron
with the electron on the neighboring site, a process which sets free
a second electron on the same sublattice. %

This picture, despite its embarrassing simplicity, can be put on firm
grounds in the context of the Valence Bond (VB) theory of chemical
bonding (see e.g. \citet{tantardini85,Cooper87,Gerratt97,McWeeny02,cooperbook02}).
Focusing on the $\pi$ electron system, this can be done with the
help of a simple (correlated) VB \emph{ansatz} for the $N$ electron
wavefunction of the $\pi$ cloud, namely\begin{equation}
\Psi_{SN}=\mathcal{A}(\phi_{1}\phi_{2}..\phi_{N}\Theta_{SN})\label{eq:VB wf}\end{equation}
where $\mathcal{A}$ is the antisymmetric projector, $\phi_{i}=\phi_{i}(\boldsymbol{r})$
for $i=1,N$ are (spatial) orbitals accommodating the $N$ electrons,
and $\Theta_{SN}$ is a $N$ electron spin-function with spin quantum
number $S$. The latter is usually variationally optimized by expansion
on a spin function basis, \[
\Theta_{SN}=\sum_{k=1,f_{S}^{N}}c_{k}\Theta_{SN;k}\]
where $f_{s}^{N}$ is the dimension of the spin-subspace of eigenfunctions
of $\mathbf{S}^{2}$ with eigenvalue $S(S+1)$ and given magnetization%
\footnote{$f_{s}^{N}$ is given by the expression $f_{s}^{N}=N!/(N/2+S+1)!/(N/2-S)!(2S+1)$
and does not depend on the value $M_{s}$ of the spin-projection $\mathbf{\hat{z}S}$
along the axis $\mathbf{\hat{z}}$, since these subspaces are isomorphic
to each other. %
}. Among these basis-functions the `perfect pairing' set devised by
Rumer, though non-orthogonal, is chemically appealing since for a
given $S$ and $M_{s}=S$ the total magnetization is given by $2S$
electrons coupled at high spin, the remaining $N-2S$ being accommodated
in $(N-2S)/2$ singlet-coupled pairs (see \citet{gianinetti68}).
Then, if the orbitals $\phi_{i}$ are \emph{localized} on the atoms,
the resulting wavefunction\[
\Psi_{SN}=\sum_{k=1,f_{S}^{N}}c_{k}\mathcal{A}(\phi_{1}\phi_{2}..\phi_{N}\Theta_{SN;k})=\sum_{k=1,f_{S}^{N}}c_{k}\Psi_{SN;k}\]
is a superposition of conventional `structures' $\Psi_{SN;k}$ describing
pairs of atom-centered, singlet-coupled orbitals (i.e. Lewis chemical
bonds and lone pairs) and unpaired electrons. `Classical' molecules
require just one perfect-pairing spin function coupling those pairs
of orbitals with substantial overlap. Less conventional molecules,
such as $\pi$ conjugated systems, need a true superposition of two
or more spin structures, since the energy gain (also known as \emph{resonance}
energy) in allowing such superposition is particularly important in
these cases. Correspondingly, the classical Lewis picture of chemical
bonds is extended to account for the resonance phenomenon, as shown
in Fig. \ref{fig:coronene-model} with double ended arrows indicating
\emph{superposition} of chemical structures. 

Early applications of the theory, starting from the landmark work
of Heitler and London, used \emph{frozen} atomic orbitals. In modern,
\emph{ab initio} use of the theory both the spin-coupling coefficients
$c_{k}$ and the orbitals can be variationally optimized, even when
using a number of configurations in place of the single orbital product
appearing in eq.(\ref{eq:VB wf}) (see e.g. \citet{McWeeny02}), in
close analogy to what is done in molecular orbital theory with the
MultiConfiguration Self-Consistent Field (MCSCF) approach. The interesting
thing is that these optimized orbitals, as a consequence of electron
correlation, are usually (if not always) localized on atomic centers
and are only slightly polarized by the environment \citep{Cooper87,CooperDavidL,Gerratt97},
thereby supporting the interpretation of the simple wavefunction of
eq.(\ref{eq:VB wf}) as a quantum-mechanical translation of Lewis
theory of chemical bond. This is true, in particular, for the benzene
molecule, the prototypical $\pi$ resonant system, where six, $p$-like
orbitals are mostly coupled by two, so-called Kekulè structures \citep{tantardini77,Cooper86,Cooper87}%
\footnote{For $S=0$ and $N=6$ the set of five linearly independent Rumer structures
is given by two Kekulè structures and the three additional `Dewar'
structures. A resonance energy $\simeq$0.8 eV can be computed when
using two Kekulè structures in place of one, whereas only some tenths
of meV are gained when full optimization of the spin function is performed
\citep{bonfanti08}. %
}. 

From a physical point of view, wavefunction (\ref{eq:VB wf}) generalizes
to $N$ electron systems the Heitler-London \emph{ansatz} forming
the basis for the Heisenberg model of magnetism in insulators. In
addition, if the orbitals are allowed to be `polarized', band-like
behavior can be accommodated, along with collective spin excitation,
as in the Hubbard model \citep{hubbard} which has been finding widespread
use in investigating graphitic compounds. The fact that Hubbard model,
and its Heisenberg limit, can be derived by suitable approximations
to Valence-Bond \emph{ansatz} has long been known in the chemical
literature, especially in connection to $\pi$ resonant systems (see
e.g. \citet{klein02,klein03} and references therein. Hubbard model
is also known as Pariser-Parr-Pople model in the chemical literature,
after Pariser and Parr \citep{pariserparr-a,pariserparr-b} and after
Pople \citep{pople53}). We can roughly say that Heisenberg models
correspond to the `classical' valence theory developed by Heitler,
London, Pauling and Van Vleck in the twenties which put the basis
for explaining chemical bond using \emph{frozen} atomic orbitals,
whereas Hubbard models arise from the modern version of theory, started
with Coulson and Fisher and pushed forward by Gerratt and others,
who used `polarizable' orbitals in the same spin scheme set up in
the original theory \citep{Cooper86,Cooper87}. 

\begin{figure}
\begin{centering}
\includegraphics[clip,width=0.8\columnwidth]{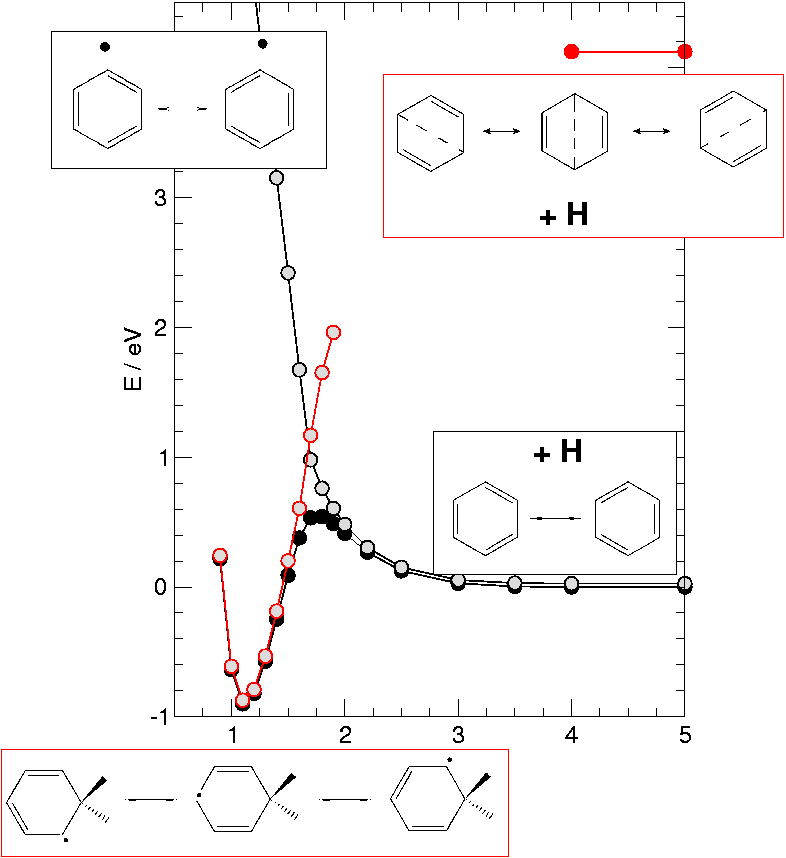}
\par\end{centering}

\caption{\label{fig:Diab}Interpretation of the sticking barrier as an avoided
crossing between chemical structures. Valence bond results for the
benzene-H system, from Ref. \citep{bonfanti08}. Solid black and red
circles for the ground ($C_{6}H_{6}({}^{1}A_{1g})+H(^{2}S)$) and
the first excited states ($C_{6}H_{6}(^{3}B_{1u})+H(^{2}S)$), as
obtained at the single-orbital-string level of eq.\ref{eq:VB wf},
with orbital optimization. Quasi-diabatic results are obtained by
properly constraining the spin space: Kelulè structures only (lower
right and upper left insets) for empty black circles; structures in
the lower left inset for empty red circles. Also shown in the upper
right inset the main (Dewar-like) structures needed to described the
$^{3}B_{1u}$ state of benzene. }

\end{figure}

In the following Sections, we will use the above wavefunction (eq.(\ref{eq:VB wf}))
as a simple guide to interpret the results of our first\emph{-}principles
calculations, keeping in mind its connections with the traditional
chemical picture on the one hand and the Hubbard model on the other.
As we will show in the following, even at this qualitative level,
a number of useful insights can be gained from such a picture. As
a first example we can reconsider adsorption of a single hydrogen
atom on graphene. In a diabatic picture (i.e. when constraining the
spin-coupling to the Kekulé structures of Fig.\ref{fig:coronene-model},
panel a) the interaction between graphene and the incoming H atom
is expected to be repulsive, since no electron is available to form
the CH bond. On the other hand, a low lying spin-excited state corresponding
to a Dewar-like structure (which has two, singlet-paired electrons
on opposite, no-overlapping end of a benzene ring) would give rise
to an attractive, barrierless interaction. At short-range, then, an
avoided crossing between the two doublet curves occurs which signals
the spin-transition leading to bond formation, even though this can
lead only to a metastable state if surface reconstruction is not allowed,
as indeed found in DFT calculations (e.g. see Fig. 2 in \citep{Sha&Jackson2002}).
Actually, in this case the situation is a bit more complicated since
a slightly lower-lying state in the triplet manifold (obtained by
spin-flipping the above spin-excited Dewar-like structure) contributes
to the same doublet manifold. Valence Bond calculations on the simpler
benzene-H system confirms this picture \citep{bonfanti08}, see Fig.\ref{fig:Diab}.

\subsection{\label{sub:Secondary-adsorption}Secondary adsorption}

\begin{figure}
\begin{centering}
\includegraphics[clip,width=0.8\columnwidth]{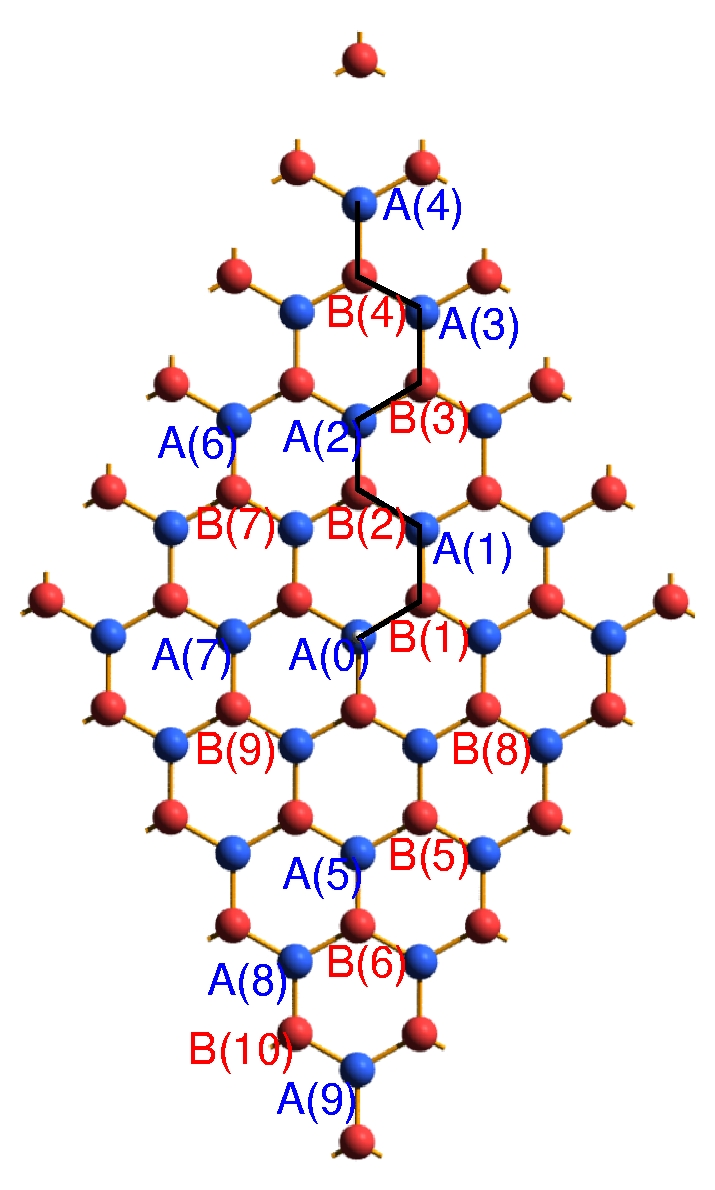}
\par\end{centering}

\caption{\label{fig: double adsorption sites}The graphene unit cell used for
the calculations with A (blue) and B (red) lattice sites indicated.
Also indicated is the path used for Fig.\ref{fig:SpinDensity}, $A(0)$
being the first H adsorption site. }

\end{figure}

Next we consider adsorption of a second atom on the different sites
A($n$) ($n=1,6$) and B($n$) ($n=1,6$) shown in Fig.\ref{fig: double adsorption sites},
with a first adsorbed H atom on site A($0$). For each site we have
investigated the ground spin manifold by allowing full relaxation
of the magnetization. In addition, in most of the cases, we have also
performed magnetization-constrained calculations in order to get insights
on both the singlet and the triplet states arising from the interaction
between the doublet H-graphene ground-state and the second H atom
\begin{table}
\begin{centering}
\begin{tabular}{|c|c|c|c|c|}
\hline 
Position & M$_{SI}$ / $\mu_{B}$ & $E$$_{\mbox{bind}}$ / eV & M/$\mu_{B}$ & $E$$_{\mbox{bind}}^{*}$ / eV\tabularnewline
\hline
\hline 
\emph{B(1)} & 0.109 & 1.934 & 0 & 0.933\tabularnewline
\hline 
\emph{A(1)} & -0.019 & 0.802 & 2 & 0.575\tabularnewline
\hline 
\emph{B(2)} & 0.085 & 1.894 & 0 & 0.828\tabularnewline
\hline 
\emph{A(2)} & -0.017 & 0.749 & 2 & 0.531\tabularnewline
\hline 
\emph{B(3)} & 0.040 & 1.338 & 0 & 0.646\tabularnewline
\hline 
\emph{A(3)} & -0.016 & 0.747 & 2 & 0.570\tabularnewline
\hline 
\emph{B(4)} & 0.076 & 1.674 & 0 & -\tabularnewline
\hline 
\emph{A(4)} & -0.016 & 0.747 & 2 & 0.573\tabularnewline
\hline 
\emph{B(5)} & 0.023 & 1.033 & 0 & 0.590\tabularnewline
\hline 
\emph{A(5)} & -0.014 & 0.749 & 2 & 0.531\tabularnewline
\hline 
\emph{B(6)} & 0.028 & 1.110 & 0 & 0.545\tabularnewline
\hline 
\emph{A(6)} & -0.015 & 0.787 & 2 & -\tabularnewline
\hline
\end{tabular}
\par\end{centering}

\caption{\label{tab:Pair energies}Binding energies ($E$$_{\mbox{bind}}$)
for secondary adsorption to form the H-pairs shown in Fig.\ref{fig: double adsorption sites},
along with the site-integrated magnetizations (M$_{SI}$) before adsorption,
and the total ground-state magnetization (M) after adsorption obtained
when fully relaxing the magnetization. Also reported the binding energies
obtained when the magnetization is constrained to $M=0,\,2\,\mu_{B}$
for A and B sites, respectively. See text for details. }

\end{table}
. 

The results for the binding energies are reported in Table \ref{tab:Pair energies},
along with the site-integrated magnetizations (M$_{SI}$) and the
total magnetization M. Site-integrated spin-densities have been obtained
by integrating the spin-density on a small cylinder (of radius half
of the C-C distance in the lattice) centered on each site, and can
be considered a rough measure of the total spin excess available on
the site. This quantity behaves very similar to the spin-density itself,
decreasing in magnitude when increasing the distance from the adsorption
site, \emph{separately} for each sublattice. Some exceptions are worth
noticing, namely the \emph{A(0)-B(5)} pair, and are due to the cumulative
effect of next-neighbors images.  Notice, however, that despite their
possible artificial nature, results corresponding to any lattice sites
when viewed \emph{as a function} of the site-integrated magnetization
give insights into the adsorption process. 

A quick look at the Table \ref{tab:Pair energies} reveals that the
two sublattices A and B behave very differently from each other, as
the spin-coupling picture of Fig. \ref{fig:coronene-model} (panels
$b$,$c$) suggests. Roughly speaking, adsorption on B lattice is
\emph{preferred} over that on the A lattice. The binding energies
are much larger than the first adsorption energy reported in Table
\ref{tab:single-atom} (they can be as large as twice the adsorption
of the first atom), and give rise to a final unmagnetized state. In
contrast, the binding energy for adsorption on a A lattice site is
comparable to that of single-H adsorption, and the ground-state of
the H-pair on graphene is a triplet (M=2 $\mu_{B}$). 

These findings agree with Lieb's theorem \citep{lieb} for the repulsive
Hubbard model of a bipartite lattice and a half-filled band, which
states that the ground-state of the system has $S=1/2|n_{A}-n_{B}|$.
In such model, the electronic state of the system would be described
by $N-2$ $p$ orbitals ($N$ being the original number of sites),
and $n_{B}=n_{A}=N/2-1$ if adsorption of the second hydrogen atom
proceeds on the B lattice (to form what we can call AB dimers), whereas
$n_{B}=n_{A}+2=N/2$ if it proceeds on the A lattice (to form A$_{2}$
dimers). The results are also consistent with the VB framework sketched
in Subsection \ref{sub:Single-atom-adsorption}: with reference to
Fig. \ref{fig:coronene-model} (panels $b$,$c$), it is clear that
when a H atom adsorbs on an B site its electron readily couples with
the unpaired electron available on the B sublattice, whereas when
adsorption occurs on an A site \emph{two} electrons are left in excess
on the B sublattice, and they more favorably couple at high spin. 

\begin{figure}
\begin{centering}
\includegraphics[clip,width=0.8\columnwidth]{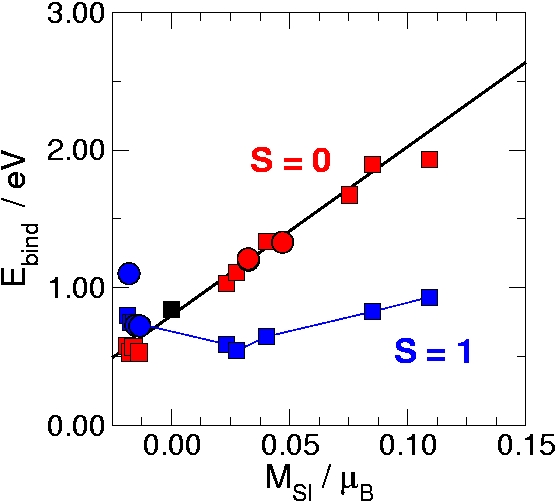}
\par\end{centering}

\caption{\label{fig:Ebind_vs_Sim}Binding energies for secondary H adsorption
as a function of the site-integrated magnetization, for singlet (red
squares) and triplet (blue squares) states. Black square is the data
point for single H adsorption. Also shown a linear fit to the data
set (solid line) and the H binding energy to form some 4-atom clusters
from 3-atom ones (red and blue circles for final singlet and triplet
states, respectively). See text for details. }

\end{figure}

The relationship between the available unpaired electron density at
a given site and the binding energy of adsorbing a second H atom can
be made clearer by reporting the energy data of Table \ref{tab:Pair energies}
as a function of M$_{SI}$. This is shown in Fig.\ref{fig:Ebind_vs_Sim},
for both the singlet and triplet states of the dimers, along with
the value for the first H adsorption (data point at M$_{SI}$=0).
It is clear from the figure that, with the exception of the value
for the \emph{ortho}-dimer ($A(0)B(1)$ in Fig.\ref{fig: double adsorption sites}.
This value has been excluded from the linear regression shown in Fig.\ref{fig:Ebind_vs_Sim}),
a linear relationship between the binding energy and the site integrated
magnetization well describes the situation, and the binding energy
for single H adsorption fits well to this picture. %

This is again consistent with the chemical model, as long as the site-integrated
magnetization is a measure of the unpaired electron density available.
According to Section \ref{sub:Single-atom-adsorption}, adsorption
of the first hydrogen atom arises from the energy balance between
a `localization energy' (the spin excitation needed to set free an
unpaired electron on the given lattice site), the spin-pairing forming
the bond, and the surface reconstruction energy. The same is true
for adsorption of a second atom: localization energy takes only a
slightly different form than before because an unpaired electron is
already available in one of the two sublattices%
\footnote{In terms of the wavefunction of eq.(\ref{eq:VB wf}) this localization
energy can be defined by observing that the structures in which the
spin-up density localizes on the $(N-1)-th$ site ($N$ even) correspond
to $\Psi=A(\phi_{1}\phi_{2}..\phi_{N-1}\Theta_{loc}^{N-1})$, where
$\Theta_{loc}^{N-1}$ is constrained to have the form $\Theta_{loc}^{N-1}=(c_{1}\Theta_{0,0}^{N-2}+c_{2}\Theta_{1,0}^{N-2})\alpha$,
whereas the ground-state spin function comprises additional contributions
from $\Theta_{1,1}^{N-2}\beta$ structures. %
}, but surface reconstruction energy is \emph{not} expected to depend
on the adsorption site. Then, adsorption energies depend on the electronic
properties only, and the linear behavior observed for singlet-state
dimers in Fig. \ref{fig:Ebind_vs_Sim} suggests that the energy needed
to localize the unpaired electron on a given site decreases \emph{linearly}
when increasing the unpaired electron density available. Notice that
negative values of M$_{SI}$ (as found at A sites), correspond to
a spin excess \emph{parallel} to that of the incoming H electron,
and for these sites localization of an unpaired electron with an \emph{antiparallel}
spin requires increasingly more energy when the (magnitude) of the
spin-density increases, since this can only be achieved by adding
one electron to the site. On the other hand, when a triplet dimer
is formed upon adsorption the H electron does \emph{not} make use
of the unpaired electron available, and adsorption energies are all
around $\sim$0.8 eV, i.e. of the order of the first H adsorption.
The effect of surface relaxation is only seen in forming the \emph{ortho}-dimer,
where few tenths of eV more than the single H relaxation energy are
required because of the closeness of the two hydrogen atoms. %
\begin{figure}
\begin{centering}
\includegraphics[clip,width=0.8\columnwidth]{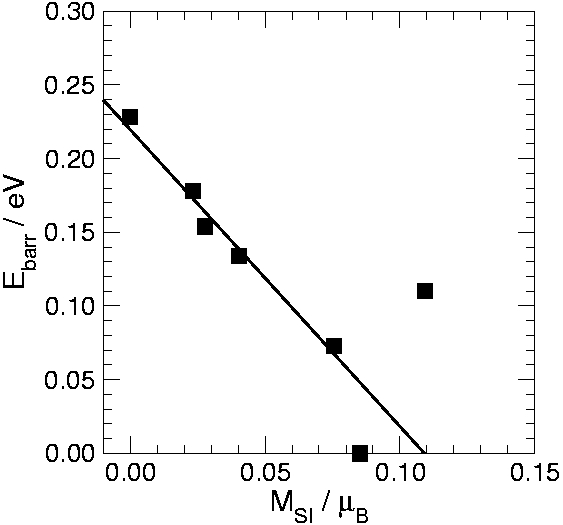}
\par\end{centering}

\caption{\label{fig:Ebarr_vs_SiM}Barrier energies for secondary atom adsorption
on the B lattice sites as a function of the site-integrated magnetization.
Linear regression of the data omits the values for forming \emph{para}
and \emph{ortho} pairs (two rightmost points in the graph). }

\end{figure}

Analogous linear behavior can be found when considering the computed
energy barrier to sticking as a function of the site-integrated magnetization,
as shown in Fig. \ref{fig:Ebarr_vs_SiM} for AB dimers. This agrees
with the above localization energy and with the common tendency for
a linear relationship between the binding and the barrier energies
for activated chemical reactions (Brønsted-Evans-Polayni rule). Exceptions
are given by the \emph{ortho-} dimer considered above and by the \emph{para-}
dimer ($A(0)B(2)$). The latter, in particular, shows no barrier to
adsorption, in agreement to previous theoretical works, and this fact
forms the basis of the so-called preferential sticking mechanism.
This mechanism was first suggested by Hoernaker \emph{et al.} \citep{Hornekaer2006a}
who looked at the STM images formed by exposing Highly Oriented Pyrrolitic
Graphite (HOPG) samples to a high-energy (1600-2000 K) H atom beam
and observed formation of stable pairs, confirmed by first\emph{-}principles
calculations \citep{Hornekaer2006a}. Our results suggest that barrierless
adsorption on the \emph{para} site is a consequence of both favorable
electronic and nuclear factors. 

We therefore find that formation of AB dimers is both thermodynamically
\emph{and} kinetically favoured over formation of A$_{2}$ dimers
and single atom adsorption. This agrees with current experimental
observations which show evidence for clustering of hydrogen atoms
at all but very low ($<1\%$) coverage conditions. In addition, we
notice that the dimers identified so far \citep{Hornekaer2006,Hornekaer2006a,Andree06}
are all of the AB type.

\subsection{\label{sub:Further-adsorptions}Further adsorptions}

We consider in this section results concerning formation of cluster
of three and four atoms. In these cases, the number of possible configurations
is quite large and therefore we limit our analysis to a few important
cases. Following analogous notation recently introduced for defects
by \citet{palacios08}, we use the `chemical formula' $A_{n}B_{m}$
to denote a cluster with $n$ H atoms in the A lattice and $m$ H
atoms in the B lattice. According to Lieb's theorem and to the $\pi$
resonance picture, we expect that the ground electronic state has
$|n-m|$ unpaired electrons. We have considered a number of $A_{2}B_{2}$,
$A_{2}B$, $A_{3}B_{1}$ and $A_{3}$ clusters, and found that their
ground-state has 0, 1, 2 and 3 $\mu_{B}$ of magnetization, respectively,
in agreement with the expectation.

Three atom clusters have been obtained by adding one hydrogen atom
either to a \emph{para} dimer or to a \emph{meta} dimer, i.e. $A(0)B(2)$
and $A(0)A(1)$ with the labels of Fig.\ref{fig: double adsorption sites},
respectively. The binding energies of a third hydrogen atom to a \emph{para}
dimer structure are reported in Tab.\ref{tab:3-atom clusters}; since
they all are of $A_{2}B$ type, the total magnetization for the resulting
structures is 1 $\mu_{B}$. A look at Tab.\ref{tab:3-atom clusters}
reveals that adsorption to a third hydrogen atom parallels that of
the first H. This is consistent with the $\pi$ resonance picture,
since $AB$ dimers do not have unpaired electrons, and therefore show
no preference towards any specific sublattice position. There are
of course exceptions, notably the values for adsorption onto $A(2)$
and $B(8)$ lattice sites, and these can be reasonably ascribed to
the effect of surface relaxation. Indeed, relaxation energies per
atom in `compact' clusters may considerably differ from the value
of the single H atom, being always of the order of the binding energies
themselves ($\sim0.8$ eV). Similar conclusions hold when adding a
third H atom to the (magnetic) \emph{meta} dimer $A(0)A(1)$: adsorption
on $B$ lattice sites is strongly favored ($E_{bind}=1.2-1.9$ eV)
and produces doublet structures ($M=1\,\mu_{B}$), whereas H atoms
bind to $A$ lattice sites with an energy $\sim0.7-0.8$ eV and produce
highly magnetic structures ($M=3\,\mu_{B}$). Energy barriers to adsorption
follow the same trend: preliminary calculations show that, with few
exceptions, barriers to sticking a third H atom compare rather well
with that for single H atom adsorption for the processes $AB\rightarrow A_{2}B$
and $A_{2}\rightarrow A_{3}$, and may be considerably smaller for
$A_{2}\rightarrow A_{2}B$ ones. %
\begin{table}
\begin{centering}
\begin{tabular}{|c|c|}
\hline 
Position & $E$$_{\mbox{bind}}$ / eV\tabularnewline
\hline
\hline 
A(2)  & 1.516\tabularnewline
\hline 
B(3)  & 0.847\tabularnewline
\hline 
A(3)  & 0.727\tabularnewline
\hline 
B(4) ($\equiv$A(5))  & 0.971\tabularnewline
\hline 
A(4) ($\equiv$B(6))  & 0.821\tabularnewline
\hline
B(7) & 0.727\tabularnewline
\hline
B(8) & 1.301\tabularnewline
\hline
\end{tabular}
\par\end{centering}

\caption{\label{tab:3-atom clusters}Binding energies ($E$$_{\mbox{bind}}$)
for addition of a third H atom to the \emph{para} dimer structure
$A(0)B(2)$ on the sites indicated in the first column (labels from
Fig. \ref{fig: double adsorption sites}). }

\end{table}

\begin{figure}
\begin{centering}
\includegraphics[width=0.9\columnwidth]{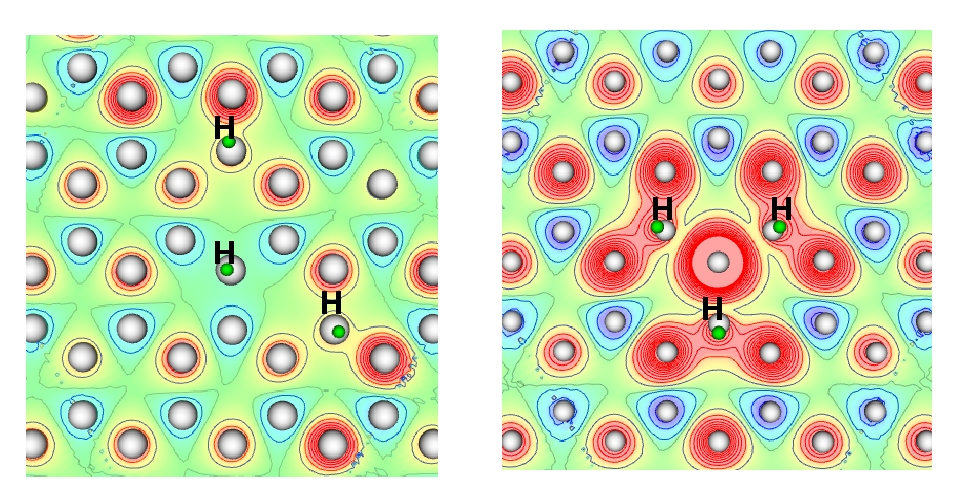}
\par\end{centering}

\caption{\label{fig:Spin-maps-3-cluster}Spin-density 0.40 \AA ~above the
surface for two three-atom clusters. Contour map with red/blue lines
for spin-up/spin-down excess respectively. Left and right panel for
an A$_{2}$B and a A$_{3}$ cluster, respectively. }

\end{figure}
In addition, again consistently with $\pi$ resonance picture, we
found that all the considered 3-atom structures, with one or two unpaired
electrons, show an alternation pattern in their spin-density maps.
As an example, Fig.\ref{fig:Spin-maps-3-cluster} reports the spin-density
maps for an $A_{2}B$ (left panel) and an $A_{3}$ (right panel) cluster.
Analogously to Subsection \ref{sub:Secondary-adsorption} we find
that analysis of these spin-density maps gives insights to the adsorption
properties of a fourth hydrogen atom. Table \ref{tab:Quad energies},
for example, reports binding energies to form some 4-atom clusters
from the stable $A(0)B(2)B(8)$ one, the final total magnetization
of the resulting structures and the values of the corresponding site-integrated
magnetization before adsorption. The computed binding energies compare
rather well with the dimer values, as can be seen in Fig. \ref{fig:Ebind_vs_Sim}
where it is clear that the results fit well to the \emph{same} linear
trend obtained before. Few exceptions are for compact clusters where
substrate relaxation does play some role. With such exceptions in
mind, our results suggest that adsorption of hydrogen atoms on magnetic
graphitic substrates (such as those obtained by adsorbing an odd number
of H atoms), for a given final spin-state, depends on the local spin-density
\emph{only}. %
\begin{table}
\begin{centering}
\begin{tabular}{|c|c|c|c|}
\hline 
 & M$_{SI}$ / $\mu_{B}$ & E$_{\mbox{bind}}$ / eV & M/$\mu_{B}$\tabularnewline
\hline 
$B(9)$ & -0.0180 & 1.103 & 2\tabularnewline
\hline 
$A(7)$ & 0.0471 & 1.331 & 0\tabularnewline
\hline 
$B(6)$ & -0.0151 & 0.727 & 2\tabularnewline
\hline 
$A(8)$ & 0.0325 & 1.210 & 0\tabularnewline
\hline 
$B(10)$ & -0.0134 & 0.723 & 2\tabularnewline
\hline 
$A(9)$ & 0.0326 & 1.201 & 0\tabularnewline
\hline
\end{tabular}
\par\end{centering}

\caption{\label{tab:Quad energies}Binding (E$_{\mbox{bind}}$) energies for
adsorption to form H-quadruples from the $A(0)B(2)B(8)$ cluster,
along with the site-integrated magnetizations (M$_{SI}$) and the
total ground-state magnetization (M), before and after adsorption,
respectively. See Fig.\ref{fig: double adsorption sites} for atom
labels. }

\end{table}

\section{\label{sec:Summary-and-Conclusions}Summary and Conclusions }

In this work we have presented results of extensive first\emph{-}principles
calculations of the adsorption properties of hydrogen atoms on graphite.
A number of possible configurations involving one, two, three and
four atoms on the surface have been considered and barrier energies
have been computed for some of them. We have found that adsorption
of hydrogen atoms is strongly related with substrate electronic properties,
and used the chemical model of planar $\pi$ conjugated systems to
rationalize the data. The connection between this model and the valence
theory of chemical bond on the one hand, and Hubbard models on the
other hand, has been emphasized in Section \ref{sub:Single-atom-adsorption},
and used at a qualitative level to rationalize our findings. In this
way, one prominent feature of defective graphitic substrates, i.e.
the possibility of forming ordered (microscopic) magnetic patterns,
turns out to be related to the spin-alternation typical of $\pi$
resonant systems. We have also invariably found in the cases considered
that Lieb's theorem for repulsive Hubbard models can be used to predict
spin alignment in ground-state graphitic structures. 

Adsorption of single H atoms has been known for some time to be an
activated process, with an energy barrier to sticking ($\sim0.2$
eV) high enough to prevent adsorption at ambient conditions. Adsorption
of a second atom more favorably occur on the $\sqrt{3}\mbox{x}\sqrt{3}R30^{o}$
sublattice where spin-density localizes, and may proceed without barrier
if it occurs on the so-called \emph{para} site. This preferential
sticking has been recently suggested by experimental and theoretical
observations (\citet{Hornekaer2006,Hornekaer2006a}). We extended
the latter analysis by considering a large number of possible dimers
and found that (i) binding (barrier) energies generally increase (decrease)
linearly as a function of the site-integrated magnetization, and (ii)
adsorption properties of the \emph{ortho} and \emph{para} sites are
slightly at variance with linear trends, thereby suggesting that substrate
relaxation plays some role in these cases. 

When considering addition of a third atom we found that the adsorption
energetics of the incoming H atom is similar to that of the first
one (i.e. a barrier $\sim0.2$ eV high and a chemisorption well $\sim0.8$
eV deep), unless we start with a `magnetic' dimer in which the two
atoms are adsorbed in the same sublattice. (These structures, however,
are kinetically and thermodynamically unfavored with respect to the
unmagnetized $AB$ configurations). This is in agreement with the
chemical model, which predicts an open-shell configuration for $A_{2}$
dimers and a closed-shell one (with partial restoring of the $\pi$
aromaticity) for $AB$ ones. These results, therefore, suggest that
preferential sticking alone cannot provide any \emph{catalytic} route
to molecular hydrogen formation on graphite. 

Finally, we have considered adsorption energetics in forming clusters
of four atoms, and re-gained the same picture obtained in forming
pairs, namely that adsorption is strongly biased towards the sublattice
in which the spin-density localizes. Actually, the resulting energetics
fits well to the linear behavior with respect to the site-integrated
magnetization already found for dimer formation. Such a linear relationship
suggests that the energy needed to localize the unpaired electron
on a given lattice site decreases linearly when increasing the site-integrated
magnetization, at least in the range of values covered by this study.
Interestingly, this behavior suggests that if we were able to tune
the magnetization of the substrate we could \emph{control} the adsorption
dynamics of H atoms. 

Overall our results, consistently with the $\pi$ resonance picture,
suggest that the \emph{thermodynamically} and \emph{kinetically} favored
structures are those that minimize sublattice imbalance, i.e. those
$A_{n}B_{m}$ structures for which $n_{I}=|n-m|$ is minimum. The
latter number $n_{I}$ is also the number of mid-gap states in single
particle spectra which, according to the Hund-like rule provided by
Lieb's theorem \citep{lieb}, is directly related to the total spin
of the system, $S=n_{I}/2$, which is therefore at minimum in the
favored structures. Notice that however small the $S$ value can be,
this result does \emph{not} preclude the existence of local magnetic
structures, antiferromagnetically coupled to each other. The case
of an $AB$ dimer with two atoms very far from each other provides
such an example.


\end{document}